\documentclass[twocolumn,preprintnumbers,amsmath,amssymb]{revtex4}

\usepackage[pdftex]{graphicx}
\usepackage{multirow}
\pdfoutput=1

\begin{document}

\title{Experimental determination of the state-dependent enhancement of
the electron-positron momentum density in solids}

\author{J.~Laverock, T.~D.~Haynes, M.~A.~Alam and S.~B.~Dugdale}
\affiliation{H.~H.~Wills Physics Laboratory, University of Bristol, Tyndall
Avenue, Bristol BS8 1TL, United Kingdom}

\begin{abstract}
The state-dependence of the enhancement of the electron-positron momentum
density is investigated for some transition and simple metals (Cr,
V, Ag and Al).  Quantitative comparison with linearized muffin-tin orbital
calculations of the corresponding quantity in the first Brillouin zone is shown
to yield a measurement of the enhancement of the $s$, $p$ and $d$ states,
independent of any parameterizations in terms of the unscreened electron
density local to
the positron. An empirical correction that can be applied to a first-principles
state-dependent model is proposed that reproduces the measured state-dependence
very well, yielding a general model for the enhancement of the
momentum distribution of positron annihilation measurements, including those of
angular correlation and coincidence Doppler broadening techniques.
\end{abstract}

\maketitle

\section{Introduction}
Positron annihilation is one of the key tools in modern investigations
of the Fermi surface (FS) of solids \cite{west1995}, alongside quantum
oscillatory techniques, Compton scattering and angle-resolved photoemission.
Unlike other methods, however, the positron probe itself plays a crucial
role in the measured distribution, preferentially annihilating with those
electrons that are most able to screen its charge. In a typical metal,
free from vacancy-type defects, the electrons that are most readily able
to screen are, of course, those at the FS, advantageously leading to an
enhancement of the signal contributed from electrons at the FS. Attempts
to account for this enhanced contribution have, for the most part, relied
on detailed studies of the electron-positron interaction within the jellium
model \cite{kahana1963,arponen1979}, which is now essentially well-understood
\cite{stachowiak1993}. However, such schemes are yet to achieve good agreement
with experiment when applied to a wide range of metallic systems. Here,
we consider this problem from an experimental perspective, {\em measuring}
the state-dependent enhancement factor for some simple elemental metals,
and present a phenomenological (and empirical) correction to the work of
Barbiellini, Alatalo and their co-workers \cite{barbiellini1997,alatalo1996}
that offers excellent agreement with experiment.

The complex many-body interaction between the positron and the electron gas has
been intensively studied for many years \cite{sormann1996}.  When the positron
enters a homogeneous electron gas, the attractive Coulomb interaction polarizes
the electron gas in the vicinity of the positron, leading to a cusp in the
unscreened electron density at the positron's position and the
associated enhancement of
the partial annihilation rate of those electrons that screen the charge.
The theory
of Kahana \cite{kahana1963} predicted a momentum-dependent enhancement,
in which the enhancement increases towards the Fermi momentum, $k_{\rm F}$,
and corresponds to the increased capability of electrons near the Fermi level
to screen the positron's charge, compared with lower-lying electron states.
However, the inhomogeneity of the electron gas in real lattices can have a
strong influence on the enhancement, even hiding the Kahana-like momentum
dependence \cite{barbiellini1997,sormann1996}.

It is worth pointing out that when considering enhancement there are
actually two separate, but related issues. Firstly, in the context of
calculating the correct positron lifetimes in solids, the enhancement of
the total electron density needs to be properly described in order to
calculate the positron annihilation rate. Secondly, a description of the
enhancement is needed when calculating the two-photon momentum densities
(which are the focus of the current paper). The former problem is easier because
the contact density can be parameterized in terms of the local electron and
positron densities (using the many-body results for jellium), but the latter
is a more difficult problem since in the framework of density functional
theory there is no formally exact way to calculate the two-photon
momentum density \cite{singh1985,barbiellini2003} (and as such all
models in the literature are, in practice, empirical).
Local density parameterizations, in which the enhancement is parameterized as
a function of the unscreened electron density, $n$, at the positron, have
been introduced
to account for the inhomogeneity of real systems. In these approaches, the
enhancement is usually expressed in terms of the electron gas parameter,
$r_s = (3/4\pi n)^{1/3}$, of which it
is a monotonically increasing function for typical crystallographic electron
densities. Some popular choices are the expressions of Arponen and
Pajanne \cite{arponen1979}, based on boson formalism and parameterized by
Barbiellini and co-workers \cite{barbiellini1995}, and those of Boro\'{n}ski
and Nieminen (BN) which are based upon an interpolation of Fermi liquid
results due to Lantto \cite{boronski1986}. Jarlborg and Singh (JS)
have used a local-density approach to solve a two-body electron-positron
Schr\"{o}dinger equation inside a spherical correlation cell that yields good
agreement with transition metals and their alloys for both momentum densities
\cite{jarlborg1987} and positron lifetimes \cite{barbiellini1991}, and is a
common choice to describe the enhancement of the momentum distribution in
metals \cite{barbiellini2003,major2004b}.  More general parameterizations
have been proposed \cite{mijnarends1979,daniuk1987,jarlborg1991,sob1982}
that include Kahana-like momentum or energy dependence to describe the
results of positron measurements.  More recently, theoretical prescriptions for
the enhancement have been developed that represent a significant departure from
the homogeneous electron gas or local-density approaches, based on, for example,
the generalized gradient approximation \cite{barbiellini1997}, Bloch-modified
ladder \cite{sormann1996} or weighted-density approximation \cite{rubaszek1998}.

Owing to the different screening properties of $d$ and $s$-$p$ electrons,
efforts to include a character, or state-dependent enhancement function
have been applied to several transition metals and their alloys
\cite{sob1982,svoboda1983,matsumoto1987,genoud1990}. \v{S}ob applied
such a scheme to data measured on a polycrystalline FeAl alloy, finding a
de-enhancement of the $d$ states by a factor of $\sim 2.2$ compared with the
$s$-$p$ states \cite{sob1982}, whereas the application of the same procedure by
Svoboda and \v{S}ob \cite{svoboda1983} to CuZn was found to favor a reduction
by a factor of $\sim 1.5$.  Theoretically, such explicit state-dependence is
rarely included, although for flat $d$-bands it is implicitly present in any
energy-dependent model. Recently, Barbiellini and co-workers have developed a
theoretical and {\em ab initio} state-dependent prescription for calculating
the enhancement in a general system \cite{barbiellini1997,alatalo1996}, which
is based on the state-dependent annihilation rates calculated within the
generalized gradient approximation (GGA). Although it has been demonstrated
that the effects of enhancement do not shift the location in {\bf k}-space of
the Fermi breaks in positron measurements \cite{majumdar1965}, the influence
of the theoretical treatment of the enhancement, when rigid-band like shifts
are applied to the electronic structure and compared with experiment, has
not yet been investigated.

Here, we tackle the problem of describing the enhancement of the
positron annihilation rates from an experimental perspective. Employing
a state-dependent (SD) model for the enhancement similar to that of Ref.\
\cite{barbiellini1997}, we simultaneously fit both the FS and the enhancement
from {\em ab initio} electronic structure calculations to positron data
directly in {\bf k}-space in order to obtain a quantitative {\em measurement}
of the enhancement in metals.  Additional comparisons with the calculational
scheme of Ref.\ \cite{barbiellini1997} are used to quantitatively assess the
applicability of such an SD enhancement model for electron-positron momentum
distributions. In particular, the accuracy of rigid-band-like approaches in
obtaining more realistic representations of the experimental FS are found
to be sensitively dependent on the particular enhancement employed in the
calculation.

The organisation of this paper is as follows. In Section \ref{s:method},
we introduce the method employed in this paper. In Section \ref{s:dmetals},
we apply this fitting technique to some $3d$ transition and noble metals,
namely V, Cr and Ag, and in Section \ref{s:alkali} we address the simple metal
Al. Finally, in Section \ref{s:model} we apply and investigate a
correction to the existing theory of Ref.~\cite{barbiellini1997} that provides
useful predictive power as a general model of enhancement in both transition
metals as well as simple metals. The application of this correction to Mo is
shown to quantitatively explain the difference in the momentum distributions of
Cr and iso-electronic Mo that is observed despite the similarity in their FS.

\section{Method}
\label{s:method}
\subsection{State-dependent enhancement}
The quantity measured by two dimensional angular correlation of
(electron-positron) annihilation radiation (2D-ACAR) experiments is a
once-integrated projection
(along a suitable crystallographic direction) of the so-called two-photon
momentum density, $\rho^{2\gamma}({\bf p})$,
\begin{equation}
\rho^{2\gamma}({\bf p}) = \sum_{i,{\bf G}} n_i
\vert C_{i,{\bf G}} \vert^2 \delta({\bf p}-{\bf k}-{\bf G}),
\label{e:rhop}
\end{equation}
where $n_i$ is the electron occupation density of state $i = \{j, {\bf k}\}$
($j$ is the band index), $C_{i,{\bf G}}$ are the coefficients of a plane-wave
expansion of the product of the electron and positron wavefunctions, in which
${\bf G}$ is a vector of the reciprocal lattice, and the $\delta$-function
expresses the conservation of crystal momentum. In a 2D-ACAR measurement, the 3D
quantity expressed in Eq.~\ref{e:rhop} is integrated along a particular
direction to yield a 2D projection of $\rho^{2\gamma}({\bf p})$, and the
projected axis is usually chosen to be a suitable high-symmetry crystallographic
axis.  The FS enters Eq.~\ref{e:rhop} as discontinuous breaks in the momentum
density when {\bf p} traverses $i$ occupied ($n_i = N$) to $i$ unoccupied ($n_i
= N-1$) (i.e. when the band crosses the Fermi energy). The folding of
crystallographically equivalent {\bf p}-points of momentum using the so-called
Lock-Crisp-West procedure \cite{lock1973} yields the `reduced momentum density'
(RMD),
$\rho^{2\gamma}({\bf k})$,
\begin{equation}
\rho^{2\gamma}({\bf k}) = \sum_{i,{\bf G}}
\vert C_{i,{\bf G}} \vert^2.
\label{e:rmd}
\end{equation}

The $C_{i,{\bf G}}$ of Eq.~\ref{e:rhop} can be written in terms of the
single-particle electron and positron wavefunctions, $\psi_i({\bf r})$ and
$\psi^+({\bf r})$ as,
\begin{equation}
C_{i,{\bf G}} = \int {\rm d}^3{\bf r}
\exp[-{\rm i}({\bf k}+{\bf G})\cdot{\bf r}] \psi_i^{\rm ep}({\bf r}, {\bf r}).
\label{e:cig}
\end{equation}
Here, $\psi_i^{\rm ep}({\bf r}, {\bf r'})$ is the electron-positron pair
wavefunction for state $i$,
\begin{equation}
\psi_i^{\rm ep}({\bf r}, {\bf r'}) = \psi_i({\bf r'}) \psi^+({\bf r})
\sqrt{\gamma_i({\bf r})},
\label{e:psiep}
\end{equation}
where $\gamma_i({\bf r})$ is the state-dependent positron enhancement
factor (for the state $i$). Setting $\gamma=1$ in Eq.~\ref{e:psiep} is
equivalent to the independent particle model (IPM), although it should be
noted that the effects of the positron wavefunction are still included in that
case. The usual parameterizations of the enhancement, for example the BN or
JS models, are based on local-density parameterizations,
in which $\gamma_i({\bf r})
= \gamma({\bf r})$ is a function only of the unscreened local electron
density at the
location of the positron. Other state-dependent prescriptions exist
(e.g.\ \cite{sob1982,jarlborg1991}), although these have relied on the {\em
empirical} determination of the state dependence of the enhancement.

Barbiellini and co-workers \cite{barbiellini1997} have proposed a theoretical
prescription for applying a state-dependent positron enhancement factor to {\em
ab initio} calculations of the electronic structure and momentum density. In
their scheme, $\gamma_i$ is obtained through the partial annihilation rates,
such that,
\begin{equation}
\gamma_i = \lambda_i / \lambda_i^{\rm IPM},
\label{e:lambdai}
\end{equation}
where $\lambda_i$ is the partial annihilation rate of the state $i$ including
correlation effects, and $\lambda_i^{\rm IPM}$ is the partial annihilation rate
due to the IPM. The total annihilation rate, $\lambda$, may be
calculated from (here shown for the local density approximation, LDA),
\begin{equation}
\lambda = {\pi}r_e^2c \int {\rm d}^3 {\bf r}\;n^+({\bf r})n({\bf r})
\gamma({\bf r}),
\label{e:lambdalda}
\end{equation}
where $r_e$ is the classical electron radius, $c$ is the speed of light and
$n^+({\bf r})$ is the positron density. In
their calculations, the GGA was used for the calculation of $\lambda_i$,
which successfully reproduces the experimental annihilation rates rather well
\cite{barbiellini1995,barbiellini1996}.

\subsection{Practical approach}
We begin by computing the {\em ab initio}
electronic structure using the linearized muffin-tin orbital (LMTO) method,
within the atomic sphere approximation and including combined correction terms
\cite{barbiellini2003}. The $C_{i,{\bf G}}$ coefficients of Eq.~\ref{e:cig} are
then computed within the IPM (equivalent to setting $\gamma=1$ in
Eq.~\ref{e:psiep}), unfolded in such a way as to resolve the individual
contribution owing to the atom index ($n$), and orbital angular momentum
quantum number ($l$),
\begin{equation}
C^{\rm IPM}_{i,{\bf G}} = \sum_{n,l} C^{\rm IPM}_{i,{\bf G},n,l}.
\end{equation}
The momentum density in the first Brillouin zone (i.e.\ the RMD,
Eq.~\ref{e:rmd}) is computed for the IPM by,
\begin{equation}
\rho^{\rm IPM}_j ({\bf k}) = {\rm constant} \times \sum_{\bf G} \vert
(\sum_{n,l} C^{\rm IPM}_{i,{\bf G},n,l}) \vert^2.
\end{equation}
For the enhancement, we introduce the quantities $\gamma_{n,l}$ that describe
the enhancement of a state of atomic species $n$ and of orbital angular momentum
$l$ ($l = s, p, d, f$).  These can then be incorporated into the calculation of
the RMD by,
\begin{equation}
\rho^{\rm SD}_j({\bf k}) = {\rm constant} \times \sum_{\bf G} \vert (\sum_{n,l}
\sqrt{\gamma_{n,l}}\;
C^{\rm IPM}_{i,{\bf G},n,l}) \vert^2.
\label{e:rhok}
\end{equation}
Note that in the above equation, the $\gamma_l$ multiply the $C_{i,{\bf G},n,l}$
coefficients, which are inside the sum over included {\bf G}-vectors, and so
the RMD must be re-computed for each $\gamma_l$ and cannot be expanded into a
sum of contributions to the momentum density from different $l$-orbitals.

In this way, $\gamma_l$ is a universal quantity, representing the partial
enhancement of a state with character $l$. The degree to which it is enhanced
depends on the coefficients of the wavefunctions in the LMTO calculation. The
enhancement, then, of a pure state of atomic species $n$ and orbital character
$l$ is given by $\gamma_{n,l} = \gamma_n \cdot \gamma_l$. Note that the band
characters (atomic species and orbital character) are strongly ${\bf
k}$-dependent, and of course vary from band to band due to hybridization with
other states, so our enhancement model is a general state-dependent model for
the enhancement (see, for example, Fig.\ \ref{f:gammak}),
but has its origins in the convenient properties of the LMTO wavefunctions.

The contribution due to core annihilations is an important consideration for
positron lifetime calculations \cite{puska1986}.  However, the core contribution
is small and relatively independent of $k$ across the first Brillouin zone (BZ),
and can safely be omitted from this
calculation.  Instead, the contribution from core states in the data is
described by a uniform background in the subsequent fitting procedure.

\subsection{Minimization procedure}
In the rigid-band
approach, the agreement between experiment and theory is iteratively
maximized with respect to a rigid shift of one or more of the energy
bands (typically those that constitute the FS), until convergence at the
minimum of the goodness-of-fit parameter is achieved.
This is similar to the method of Ref.~\cite{major2004b}, however, there
are some important differences.
In Ref.~\cite{major2004b}, the radial anisotropy of the
two-photon momentum density in {\bf p}-space served as the comparative
quantity, and the enhancement
was fixed to that chosen in its initial calculation (in that case, the JS model
was used). Here, we perform our comparison in {\bf k}-space, corresponding to
the Lock-Crisp-West-folded data, and explicitly include enhancement of the form
outlined above (SD model) in the fitting.
The advantage of operating in {\bf k}-space (aside from the 
smaller array sizes involved) is
principally that we are sensitive directly to the projected Fermi breaks, rather
than the many weaker FS signatures that are distributed throughout the {\bf
p}-space spectrum.  An additional consideration, however,  is the contribution
from higher momentum components (Umklapp processes), whose enhancement has
presented a challenge for theoretical models (see, for example, 
Ref.~\cite{sormann1996}).  It is noted that operating in {\bf k}-space
involves the
folding of the Umklapp contributions into the first BZ, both experimentally and
theoretically, and that any non-trivial behavior of these contributions is
subsequently lost.
However, we have checked our results with the equivalent {\bf p}-space spectra,
and in particular near the Umklapp regions (as well as its integral, which
represents an analogue of the coincidence Doppler broadening spectra).
Crucially, we find the data are equally well-reproduced using such a
{\bf k}-space approach as they are with the traditional JS model.

The fitting parameters constitute the energy shift, $\delta_j$, for each band in
the fit (typically those that cross $E_{\rm F}$), two scaling parameters for
each experimental projection, $\Delta_m$ and $S_m$, that approximately relate to
the ${\bf k}$-independent core contribution and the number of counts in the
2D-ACAR spectra respectively, and the enhancement parameters, $\gamma_{n,l}$, of
which there are typically three for simple systems.
These are simultaneously adjusted using the {\tt MINUIT} package
\cite{minuit} and the computed {\bf k}-space density is compared with the data
until convergence is reached.
Note that we fit the ratios of the enhancement
parameters, absorbing their magnitude into the scaling parameters (the absolute
magnitude of the enhancement parameters is indistinguishable from the scaling
parameters in the data).

Our definition and treatment of the scaling parameters have an important
consequence. As mentioned above, we do not treat the enhancement of the core
electrons, preferring to concentrate on the description of the shape of the RMD.
Such an approach means that good agreement can be obtained with the IPM if we
consider a {\em negative} contribution from core annihilations. Whilst this is
clearly unphysical, it stems from the strong overestimation of the enhancement
of deeply-bound electrons within the IPM. Here, we are most interested in the
band properties of the momentum distribution (i.e.\ its {\em shape}),
and in particular its FS signatures. We point out that in the following
discussion, even when the IPM appears to give reasonable agreement with our
data, the agreement with positron lifetime measurements (see, for example,
Refs.~\cite{barbiellini1991,puska1986,takenaka2008}) would be very poor,
in contrast to the other enhancement models that are addressed here.

\section{Transition and noble metals}
\label{s:dmetals}
The transition metals and their alloys have traditionally been the subject of
the bulk of experimental investigations of the FS, and a good description of the
electron-positron momentum density and enhancement of such systems has been
vital in understanding 2D-ACAR, and indeed coincidence Doppler broadening
\cite{alatalo1996,asokakumar1996,makkonen2006} data.  The JS model was
specifically developed with transition metals in mind, and is generally thought
to provide a good description of the enhancement for $d$-electron densities
(with $r_s \sim 1.8$) \cite{jarlborg1987,barbiellini1991,major2004b}. Here, we
begin by applying the SD enhancement model described above to
some metals whose FSs have been accurately determined via quantum-oscillatory
methods (Ag, V) and one whose FS is inaccessible to conventional FS
probes (paramagnetic Cr), first concentrating on the ``raw'' LMTO
calculations of the RMD. Comparisons are made with both the IPM and the JS model
for enhancement, as well as a simplified version of the Barbiellini-Alatalo
\cite{barbiellini1997,alatalo1996} enhancement scheme. Following this, we
rigidly fit the bands to the experimental data to obtain an experimental
measurement of the FS, in order to assess the sensitivity of this approach to
the FS.

\begin{figure}[t]
\begin{center}
\includegraphics[width=1.0\linewidth]{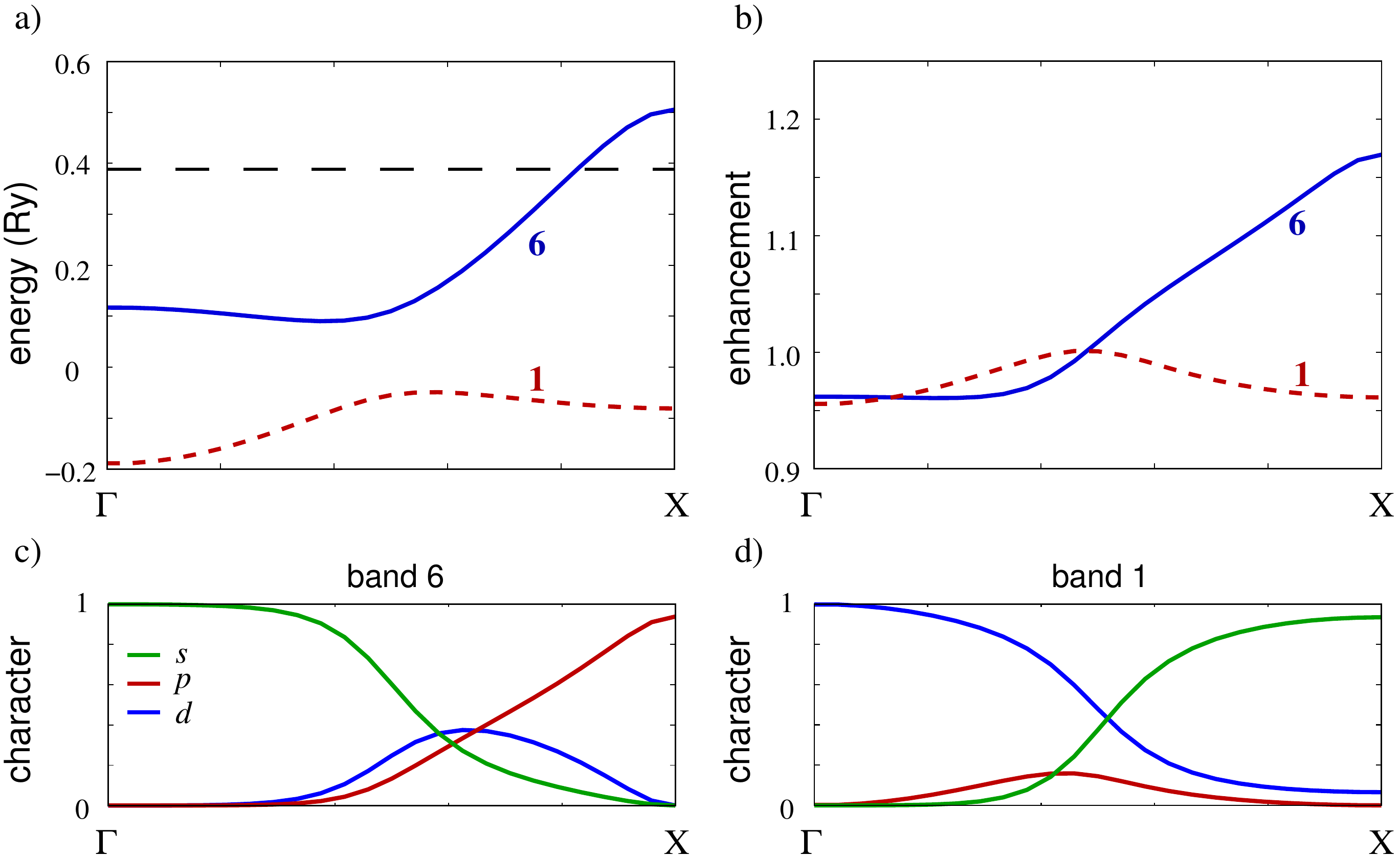}
\end{center}
\caption{(color online) The state-dependent enhancement of Ag from our model,
shown for two
energy bands along the path $\Gamma$-$X$ in the BZ.
(a) The dispersion of bands 1 and 6; (b) the enhancement from our
fit to the experimental data; (c) and (d) the character of bands 6 and 1
respectively.
Note that band 6 crosses $E_{\rm F}$, shown by the dotted
line in (a); above which the enhancement is unphysical.}
\label{f:gammak}
\end{figure}

Three 2D-ACAR projections ([100], [110] and [111]) were obtained from a single
crystal of Ag at $\sim 70$~K, with a resolution full width at half maximum
of $0.71 \times 1.11$ mrad in the $p_x$
and $p_y$ data axes respectively (corresponding to $\sim 12 \% \times 19$ \% of
the BZ of Ag). For V, four projections were obtained along the [100],
[110], [210] and [211] directions at room temperature with a resolution of
$1.11 \times 1.33$ mrad (with the exception of the [110] direction, which was
collected at $\sim 24$ K with resolution $0.83 \times 1.11$ mrad). Paramagnetic
Cr was measured along the [100] and [110] directions at 353~K, well above
the N\'{e}el temperature (with
a resolution function the same as the room temperature V measurements).
For each sample presented in this manuscript, the 2D-ACAR spectra have been
carefully checked to confirm the absence of any defect or impurity signatures
in the spectra.

\begin{figure*}[t]
\begin{center}
\includegraphics[width=1.0\linewidth]{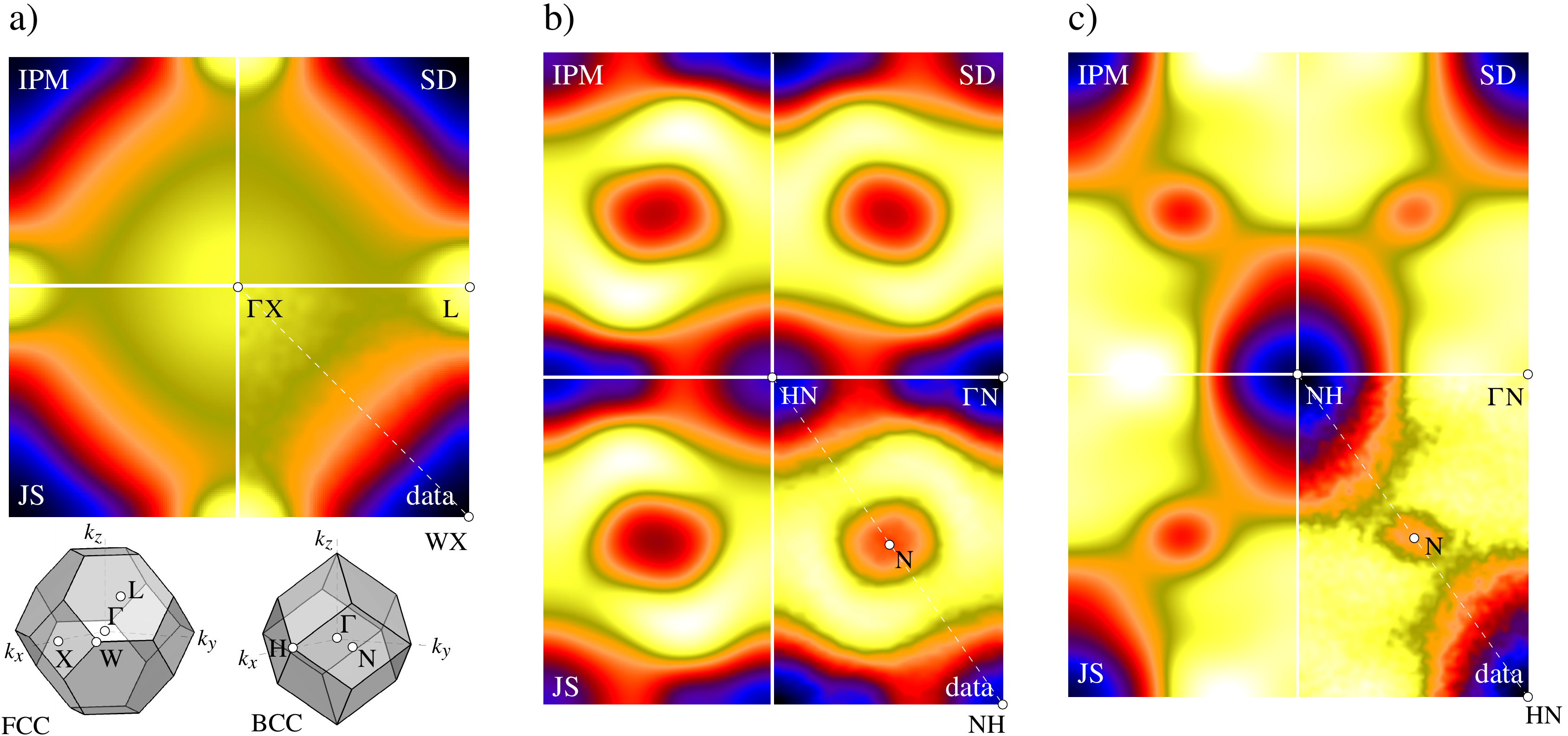}
\end{center}
\caption{(color online) Comparison between experimental {\bf k}-space momentum
density for (a) Ag [100] projection,
(b) V [110] projection and (c) Cr [110] projection
and the computed raw band calculation of the RMD for IPM, JS and SD models. High
symmetry points in projection have been labelled.
Note that in (a) the horizontal and vertical axes are $\left<110\right>$
crystallographic axes; the $\left<100\right>$ axes are along the diagonal.}
\label{f:rawdata}
\end{figure*}

LMTO calculations were performed over 1505 k-points in the irreducible wedge of
the face-centered cubic BZ for Ag, and over 6201 k-points in the irreducible
wedge of the body-centered cubic BZ for V and Cr. For each material, the RMD was
computed for both the IPM and the JS parameterizations of the enhancement,
and convoluted with the appropriate experimental resolution function. This was
compared to the experimental data with adjustments only to the scaling
parameters, $S_m$ and $\Delta_m$.  The SD model of the enhancement was then
obtained by simultaneously fitting the orbital enhancement factors, $\gamma_l$.
In each
case, a goodness-of-fit parameter, $\chi_{\rm red}^2$, was computed as a
weighted average of that from each experimental and theoretical projection.
Comparisons were also made for the raw band calculation with the model of Refs.\
\cite{barbiellini1997,alatalo1996} by computing the annihilation rates
associated with each orbital both within the GGA \cite{barbiellini1995} and IPM,
where the enhancement of each orbital in this model is the ratio of these
annihilation rates. However, it should be noted that this scheme still invokes a
parameterization of the enhancement in terms of the electron-gas parameter. To
compare with our experimental values, we integrate over {\em all} ${\bf
k}$-states to obtain the annihilation rate from all electrons of orbital quantum
number $l$. In Eq.~\ref{e:lambdalda}, this corresponds to substituting the
(partial) electron density due to each $s$, $p$, $d$ and $f$ orbital, $n_l({\bf
r})$ for $n({\bf r})$ to obtain $\lambda_l$ in the GGA, rather than the method
of Ref.\ \cite{barbiellini1997} in which $n_{j, {\bf k}}({\bf r})$ is used. In
the following, we refer to this as the simplified Barbiellini-Alatalo (SBA)
model (in which the `simplified' reflects the integration over all {\bf
k}-states).

\subsection{Raw band calculations}
\begin{figure}[t]
\begin{center}
\includegraphics[width=1.0\linewidth]{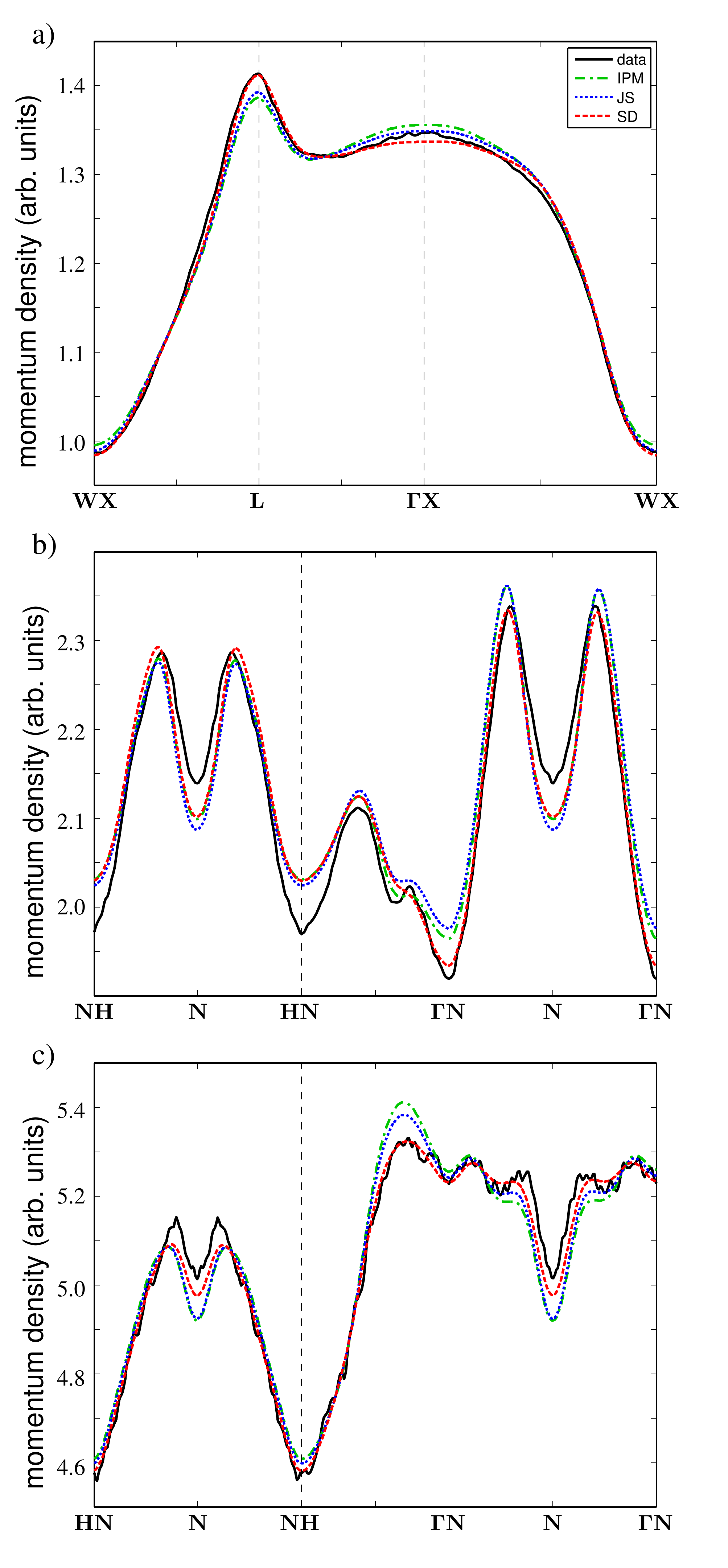}
\end{center}
\caption{(color online) The raw-band RMD of (a) Ag, (b) V and (c) Cr shown in
Fig.\ \ref{f:rawdata}, shown here along a path in the BZ.}
\label{f:rawline}
\end{figure}

For the raw band calculations, the comparison between theory and experiment
depend on (i) a good {\em ab initio} description of the electronic structure,
and in particular the FS, and (ii) a reliable understanding of the
electron-positron enhancement factor. The well-known FS of Ag consists of just a
single sheet that is only slightly perturbed from the free-electron sphere, most
notably along the [111] direction where the FS intersects with the BZ boundary
to form a neck at the $L$-point of the BZ. Band structure calculations within
the LDA reproduce the precise measurements of quantum oscillations
\cite{halse1969,coleridge1972,coleridge1982} very well and it therefore provides
an excellent candidate in which to test models for the enhancement, for which
the JS model would be an obvious choice.  As demonstrated by
Fig.~\ref{f:rawdata}a and \ref{f:rawline}a, the experimental data are
well-described by the raw band calculations of the RMD, in which even the IPM
(including only positron wavefunction effects) works reasonably well.
Quantitatively, as demonstrated in Table \ref{t:rawband}, the JS enhancement
model is found to improve the agreement between experiment and theory,
particularly near the projected $\Gamma{X}$ and $L$ points of the BZ.  However,
the current SD model is able to bring the theoretical RMD into much closer
agreement with the data by de-enhancing the $s$ and $d$ states relative to the
$p$ states.  For Ag, the bands below $E_{\rm F}$ are predominantly $d$-bands,
with some hybridization with the $5s$ state, but the band that crosses $E_{\rm
F}$ has substantial $p$ character. The de-enhancement of $d$ states relative to
the $sp$ bands is well-known \cite{jarlborg1987} and is attributed to the
relative localization of $d$ electrons, particularly near the top of the
$d$-bands. That the $p$ enhancement appears to be quite strong can be explained
by a Kahana-like momentum enhancement, in which those electron states nearest
${\bf k}_{\rm F}$ are most enhanced. In Fig.\ \ref{f:gammak} the measured SD
enhancement is plotted along $\Gamma$-$X$, accompanied by the band dispersion
and character. As can be seen from Fig.\ \ref{f:gammak}b, the enhancement of
band 6 grows substantially as the band approaches the Fermi level, replicating
the Kahana-like momentum dependence of the enhancement. This is captured in our
model by the enhancement of the $p$-like states; as demonstrated by Fig.\
\ref{f:gammak}c, the enhancement of both bands 1 and 6 closely follow their
respective $p$ character.
Note that, owing to the weak contribution from high-lying $f$ states, a good
quantification of their enhancement is not possible.

\begin{table}[tb]
\begin{center}
\begin{tabular}{||c|c||cccc||c||}
\hline
 \multicolumn{2}{||c||}{} &
 $\gamma_s$ & $\gamma_p$ & $\gamma_d$ & $\gamma_f$ & $\chi_{\rm red}^2$ \\
\hline
\multirow{4}{*}{\bf Ag}   & IPM &  -   &  -   &  -   &   -    & 12.51 \\
                              & JS  &  -   &  -   &  -   &   -    & 10.37 \\
                              & SD  & 0.81 & 1.00 & 0.81 & (0.76) &  6.57 \\
                              & SBA & 0.92 & 1.00 & 0.64 & (1.08) &  8.98 \\
\hline
\multirow{4}{*}{\bf V} & IPM &  -   &  -   &  -   &   -    & 19.74 \\
                              & JS  &  -   &  -   &  -   &   -    & 26.56 \\
                              & SD  & 0.69 & 1.00 & 0.78 & (0.57) & 13.32 \\
                              & SBA & 0.97 & 1.00 & 0.83 & (1.02) & 32.86 \\
\hline
\multirow{4}{*}{\bf Cr} & IPM &  -   &  -   &  -   &   -    &  8.15 \\
                              & JS  &  -   &  -   &  -   &   -    &  5.82 \\
                              & SD  & 0.82 & 1.00 & 0.54 & (0.87) &  2.54 \\
                              & SBA & 0.97 & 1.00 & 0.80 & (1.03) &  5.09 \\
\hline
\end{tabular}
\end{center}
\caption{The results of the fit between the different parameterizations of the
enhancement and the data.
For the SD
model, the $\gamma_l$ for each state obtained from the fit is also given,
normalized to $\gamma_p = 1$. The $\gamma_l$ for the SBA model are determined
from the partial annihilation rates described in Eq.\ \ref{e:lambdai}. The
errors in the fit of the $\gamma_l$ of the SD model are $\sim \pm 0.01$. Note
that the higher statistical precision of the V data yields a relatively large
$\chi_{\rm red}^2$ parameter.}
\label{t:rawband}
\end{table}

The situation is more complicated for V and Cr, for which the details of the
near-$E_{\rm F}$ electronic structure, including the precise dimensions of the
FS, are either not well-reproduced by our band calculations (V), or have not
been accurately determined (paramagnetic Cr). Figs.\ \ref{f:rawdata}b and
\ref{f:rawdata}c show representative 2D-ACAR projections of V and Cr (along the
[110] direction) respectively, compared with the corresponding theoretical
quantities. The FS of V is composed of two sheets.  The first sheet (originating
from band 2 of predominantly $3d$ character) forms a small $\Gamma$-centred hole
octahedron that encloses $\sim 0.12$ holes and has remained unobserved in
quantum oscillation data, although its presence has been confirmed by 2D-ACAR
measurements \cite{singh1985b,pecora1988,singh1984} (this sheet is visible in
the experimental data in Fig.~\ref{f:rawdata}b at the projected ${\Gamma}N$
point).  Band 3, on the other hand,
experiences appreciable hybridization with the $4p$ states above $E_{\rm F}$ and
forms a $\Gamma$-centred jungle-gym hole FS as well as some hole ellipsoids that
are centred at $N$. These $N$-hole ellipsoids can be clearly seen in the data of
Fig.\ \ref{f:rawdata}b at the projected $N$-point of the BZ, where the density
experiences a local dip due to their presence. These features in the IPM and JS
calculations are predicted to be substantially too large and too strong (see
Fig.\ \ref{f:rawline}b), and the enhancement of the high-density surrounding
region is not well reproduced. Although, as pointed out by Jarlborg and Singh,
the enhancement is expected to be less important for a less-full $d$-band, the
JS enhancement is actually found to perform worse than the IPM for V (see Table
\ref{t:rawband}), at least in the shape of the distribution (positron lifetime
predictions are substantially better described by the JS model
\cite{barbiellini1991}).
V was used as a test material by Jarlborg and Singh in their
presentation of the JS enhancement model, in which they comment that the IPM
already provides a reasonable description of the momentum density
\cite{jarlborg1987}, and that their model offered only weak improvement.
However, their comparisons were made with electronic structure calculations
where the bands had been rigidly shifted to agree with de Haas-van Alphen (dHvA)
measurements of the $N$-hole ellipsoids \cite{parker1974}. Indeed, in Fig.\
\ref{f:rawline}b it is obvious that the dimensions of the dips in the momentum
density along $NH$-$N$-$HN$ are incorrectly placed with respect to the data. In
Section \ref{ss:rigidband}, we address such inconsistencies by rigidly shifting
the bands to improve agreement between experiment and theory.

Cr neighbors V in the periodic table, having an extra electron,
yet its paramagnetic FS has remained relatively unexplored experimentally,
principally owing
to the emergence of an ordered spin-density wave phase below $\sim 312$ K
\cite{fawcett1988}, where the high temperature precludes quantum oscillatory
measurements in the paramagnetic phase, and strong spin fluctuations appear to
suppress the measurement of the nested sheets in the ordered phase
\cite{ditusa2010}. Theoretically,
the FS is composed of three sheets, the first of which (band 3) contributes some
small electron `lenses' midway between $\Gamma$ and $H$. Band 4 forms some $N$
hole ellipsoids and $H$-centred octahedra, whereas in band 5 there is a
$\Gamma$-centred electron `jack'. Molybdenum, isoelectronic to Cr, shares
a similar FS topology, in which the $N$-hole ellipsoids and the electron jack
can be clearly visualized in the [110]-projected {\bf k}-space density of
2D-ACAR measurements \cite{fretwell1995,hughes2004}. For Cr, these features,
shown in Fig.\ \ref{f:rawdata}c near the projected $N$ points, are obscured
in the measurement, presumably owing to
enhancement effects \cite{dugdale1998}.  Indeed, the $N$-hole ellipsoids are
more evident in the IPM and JS projected densities than they are in the data.
Overall, the agreement between the IPM and JS calculations of the RMD and the
data is reasonable (see Table \ref{t:rawband}), but is particularly poor near
the projected $N$-points of the BZ as well as midway along the $NH$-$\Gamma{N}$
path (see Fig.\ \ref{f:rawline}c). Here, the knobs of the electron jack project
on top of one another, and the IPM and JS do not predict the de-enhancement of
the momentum density very well in this part of the BZ.

The application of the SD model, however, considerably improves the agreement
between experiment and theory by substantially de-enhancing the $s$ and $d$
states. In Figs.\ \ref{f:rawline}b and \ref{f:rawline}c, this can be most
clearly seen at the projected $N$-points of the BZ, as well as the momentum
density near the ${\Gamma}N$ points. For both V and Cr, the hybridization of the
valence states with the unoccupied $4p$ states indicates the importance of the
$p$ electrons in deciding the topology of the FS, and their proximity to $E_{\rm
F}$ means they have a strong impact on the enhancement of the momentum density
in 2D-ACAR measurements. Previous non-iterative comparisons of orbital-weighted
band theory and 2D-ACAR data have been made in {\bf p}-space for V
\cite{genoud1990}, in which a de-enhancement of the $s$ and $d$ states by $\sim
0.8$ relative to the $p$ states was favored. Our results are close to these,
where we obtain 0.69 and 0.78 for $s$ and $d$ states respectively, corresponding
to a slightly greater de-enhancement of the lower-lying $s$ states. Similarly
for paramagnetic Cr, Matsumoto and Wakoh \cite{matsumoto1987} estimated
(also non-iteratively) that the Cr $d$ states were de-enhanced by $\sim 0.67$
relative to the $sp$ states (which were considered together). Our results
correspond well with their findings, where we obtain 0.82 and 0.54 for $s$ and
$d$ states, relative to the $p$ states. The stronger enhancement of the Cr $s$
states (compared with V) may be explained by the higher occupation of the $4s$
states in Cr (they are almost twice as occupied in Cr).

Finally, we comment on the predictions for state enhancement made by the SBA
model. Apart from Ag, which has a much higher $d$ electron density than either V
or Cr, the orbital enhancement ratios are predicted to be very similar (see
Table \ref{t:rawband}), with a weak de-enhancement of the $s$-states and a
modest de-enhancement of the $d$-states relative to the $p$-states. For Ag, a
rather more exaggerated de-enhancement is predicted for the $d$-states.
Qualitatively, these results are in agreement with our measured values, but
differ substantially in magnitude and lead to a slightly higher $\chi^2_{\rm
red}$ parameter than the current SD model. Nevertheless, the SBA model provides
better agreement with the data than either the IPM or the JS model for Ag and
Cr, supplying a more robust predictive scheme for computing the enhancement in
2D-ACAR momentum distributions.
That it does not fair so well for V is mostly accounted for by the rather
larger corrections to the LDA band structure that are required for V, a
topic that will be returned to in the next section. Here, we emphasize that
the SBA model is expected to improve the agreement with the data over the
IPM or JS models when extensions to the LDA, such as non-local potentials
\cite{barbiellini1989} or self-energy corrections \cite{barbiellini2005},
that improve the description of the FS are included.
The most probable origin for the discrepancy between the SBA
scheme and our measured SD enhancement is the omission of a Kahana-like
energy-dependence or momentum-dependence in the predictive scheme of
Ref.~\cite{barbiellini1997}. As already highlighted, some of the results for the
enhancement in the current SD model, particularly the apparent strong $p$
enhancement in V and Cr, reflect a Kahana-like enhancement of those electron
states near $E_{\rm F}$. In our model, where a state is not too dispersive in
energy, this is naturally captured by enhancement of that state.  The authors of
Ref.~\cite{barbiellini1997} comment that the Kahana-like energy- or
momentum-dependence appears to be less important than the state-dependence from
their results, a conclusion that this work substantiates, but these results
suggest that including such enhancement could produce a good improvement in the
agreement between experiment and theory.  In Section \ref{s:model}, we apply
just such an energy-dependent term to the SBA enhancement factor, and
demonstrate the improved predictive capacity of such a model.

\begin{figure}[t]
\begin{center}
\includegraphics[width=0.73\linewidth]{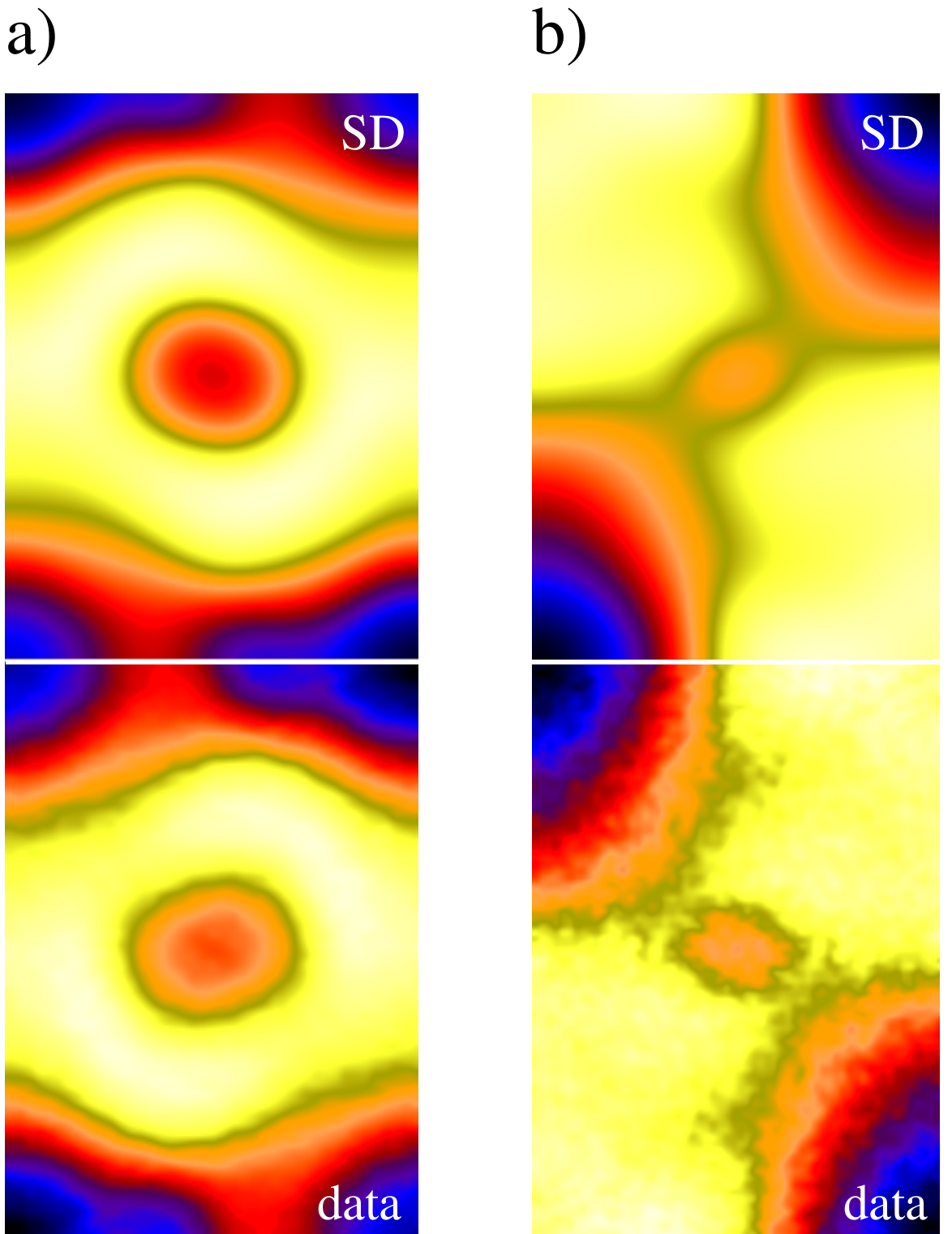}
\end{center}
\caption{(color online) Comparison between experimental data for
(a) V [110] projection and (b) Cr [110] projection
and the rigid band fit to the RMD for the SD model, shown in the same
way as Fig.\ \ref{f:rawdata}.}
\label{f:fitdata}
\end{figure}

\begin{figure}[t]
\begin{center}
\includegraphics[width=1.00\linewidth]{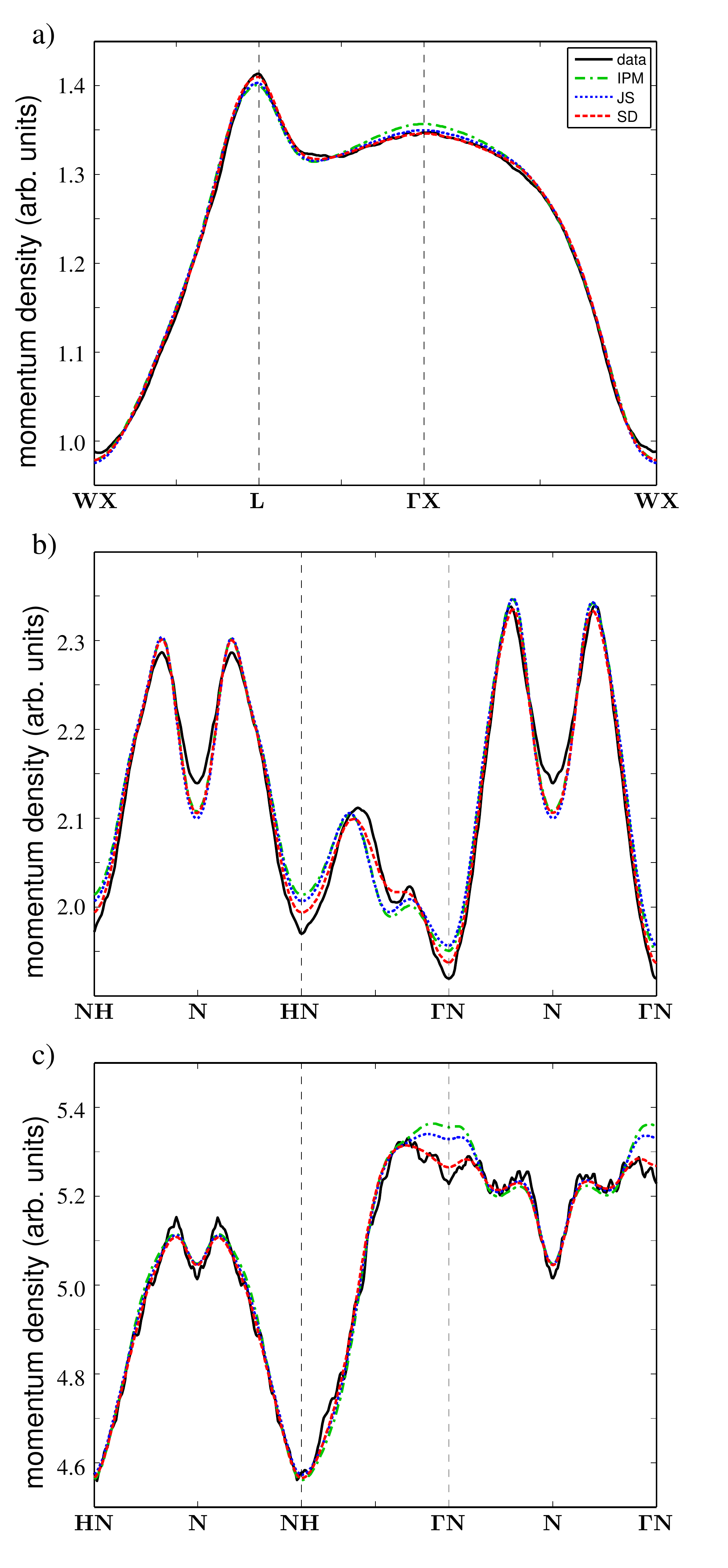}
\end{center}
\vspace*{-0.2in}
\caption{(color online) The rigid-band fit to the RMD of (a) Ag, (b) V and (c)
Cr shown in Fig.\ \ref{f:fitdata}, shown here along a path in the BZ.}
\label{f:fitline}
\end{figure}

\subsection{Rigid-band fit}
\label{ss:rigidband}
In 2D-ACAR investigations of the FS, the traditional method of extracting the FS
from experimental data is to contour the data at a level that corresponds to
extrema in the first derivative of the data, and it is well known that
enhancement effects do not shift the location of these breaks
\cite{majumdar1965}.
While first-principles calculations are often able to make excellent qualitative
predictions about the nature of the Fermi surface, when subject to detailed
scrutiny in light of precise experimental data it is often found that
quantitative differences exist. Shortcomings in the approximations used in the
calculations (e.g.\ exchange-correlation functional, neglect of relativistic
effects) mean that in reality it is difficult to get the Fermi surface correct.
These differences can often be reduced or eliminated by small shifts of the
relevant bands with respect to the Fermi level, and
it has recently become feasible, and indeed quite common, to `tune' a
band-theoretical calculation in this way
(e.g. see Refs.~\cite{major2004b,utfeld2009,utfeld2010}).
Such an approach requires an accurate
description of the positron enhancement, if conclusions regarding details of the
FS itself are to be drawn from such a fit, and we now turn our attention to
investigate the behavior of our SD model applied to such detailed FS studies.

For the rigid-band fit of the electronic structure, the Fermi level for
each band near $E_{\rm F}$ was fitted to the data. In the case of the SD
enhancement model, the orbital enhancement factors were fitted simultaneously.
The $\chi_{\rm red}^2$ was computed as before, and the number of electrons
enclosed by the fitted FS (i.e. the occupied fraction of the Brillouin zone)
was obtained.

The results of the rigid-band fit to the data are displayed in Fig.\
\ref{f:fitdata}, and demonstrate substantial improvement over the corresponding
raw band calculations of Fig.\ \ref{f:rawdata}. In Table \ref{t:rigidfit},
the $\chi^2_{\rm red}$ is shown for each fit, along with the fitted orbital
enhancement factors for the SD model. Beginning with some general comments,
we note that the orbital enhancement parameters obtained in the
SD model are only moderately adapted as a consequence of including the
bands in the fit, and the same general trends are observed.  Additionally,
it is noteworthy that in almost every case the shift in the energy band
(see Table \ref{t:bandshifts}) is found to be smallest for the SD model
(with the exception of band 3 of V).

Fig.~\ref{f:fitline} shows the RMD along the same path through the BZ 
as in Fig.~\ref{f:rawline}, and we will now concentrate in more detail on the
agreement between experiment and theory.
As indicated by the small change
in electron count of the shifted bands, the change in the FS itself is small,
owing to the appreciable dispersion of band 6 at $E_{\rm F}$, and the shift
in the Fermi wavevector is ${\Delta}k_{\rm F} \sim 0.02\,(2\pi/a)$
(just $\sim 15\%$ of the resolution function).

\begin{table}[tb]
\begin{center}
\begin{tabular}{||c|c||cccc||c||}
\hline
 \multicolumn{2}{||c||}{} &
 $\gamma_s$ & $\gamma_p$ & $\gamma_d$ & $\gamma_f$ & $\chi_{\rm red}^2$ \\
\hline
\multirow{3}{*}{\bf Ag}   & IPM &  -   &  -   &  -   &   -    &  5.53 \\
                              & JS  &  -   &  -   &  -   &   -    &  4.84 \\
                              & SD  & 0.88 & 1.00 & 0.85 & (0.76) &  4.09 \\
\hline
\multirow{3}{*}{\bf V} & IPM &  -   &  -   &  -   &   -    &  7.77 \\
                              & JS  &  -   &  -   &  -   &   -    &  8.40 \\
                              & SD  & 0.61 & 1.00 & 0.63 & (0.49) &  3.79 \\
\hline
\multirow{3}{*}{\bf Cr} & IPM &  -   &  -   &  -   &   -    &  2.52 \\
                              & JS  &  -   &  -   &  -   &   -    &  1.85 \\
                              & SD  & 0.83 & 1.00 & 0.61 & (1.27) &  1.28 \\
\hline
\end{tabular}
\end{center}
\caption{The results of the rigid-band fit between the different
parameterizations of the enhancement and the data, presented in the same
way as Table \ref{t:rawband}. The shifts in the energy bands for each fit
are shown in Table \ref{t:bandshifts}.}
\label{t:rigidfit}
\end{table}

The improvement is much more dramatic for V (Fig.\ \ref{f:fitline}b),
in which the size of the $N$-hole features in the data is now well-described
by all of the enhancement models, stemming from opposite shifts in bands 2
and 3.
After rigidly shifting the bands, the IPM and JS demonstrate similar shifts of
the energy bands and a similar goodness-of-fit parameter, leading to an excess
in occupied volume of 0.14 and 0.10 of an electron
The improvement in the SD model is more pronounced, however.
The $d$ bands are well known
to be placed too low by the LDA with respect to $sp$ bands. As noted in
Ref.~\cite{major2004b}, band 2 (of predominantly $d$ character) is pushed up in
energy by the fit towards the higher $p$ character of band 3, which is pulled
down by the fit, correcting this tendency.

\begin{table}[t]
\begin{center}
\begin{tabular}{||c|c||ccc||c||}
\hline
\multicolumn{2}{||c||}{} & \multicolumn{3}{c||}{band shifts / mRy} &
electron +/- \\
\hline
\multirow{4}{*}{\bf Ag}   &     & band 6 &   -    &   -    &      \\
\cline{2-6}
                              & IPM & -21.2  &   -    &   -    &+0.08 \\
                              & JS  & -18.3  &   -    &   -    &+0.07 \\
                              & SD  & -14.2  &   -    &   -    &+0.05 \\
\hline
\multirow{4}{*}{\bf V} &     & band 2 & band 3 &   -    &      \\
\cline{2-6}
                              & IPM & +22.3  & -15.8  &   -    &+0.14 \\
                              & JS  & +26.5  & -15.4  &   -    &+0.10 \\
                              & SD  & +18.8  & -16.4  &   -    &+0.17 \\
\hline
\multirow{4}{*}{\bf Cr} &     & band 3 & band 4 & band 5 &      \\
\cline{2-6}
                              & IPM & -25.1  & +15.6  & +13.2  &-0.02 \\
                              & JS  & -22.9  & +12.2  & +13.2  &-0.01 \\
                              & SD  & -18.5  &  +5.0  & +13.2  & 0.02 \\
\hline
\end{tabular}
\end{center}
\caption{The shifts in the energy bands for each of the rigid-band fits. Also
shown is the change in electron count in the BZ due to the fit. Note that for
Cr band 5 (that just grazes $E_{\rm F}$)
is completely expelled by all of the fits. The errors in the shifts
of the bands are in each case $\lesssim 1$~mRy.}
\label{t:bandshifts}
\end{table}

\begin{figure}[t]
\begin{center}
\includegraphics[width=1.0\linewidth]{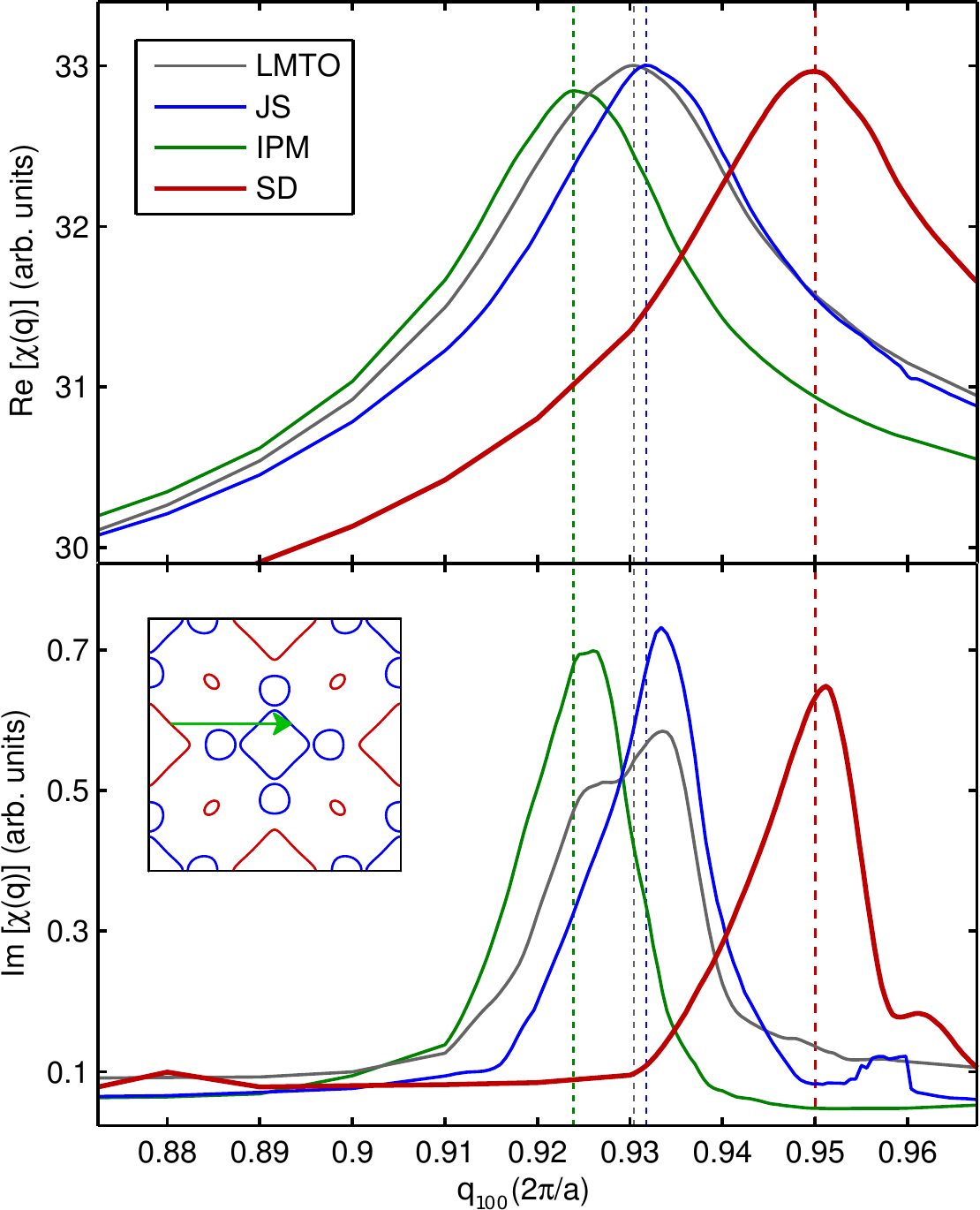}
\end{center}
\caption{(color online) The real (top)
and imaginary (bottom) parts
of the static susceptibility,
$\chi_0({\bf q})$, of paramagnetic Cr, calculated for the raw
band LMTO calculations as well as the results of the rigid-band fit to the data
with the IPM, JS and SD models of enhancement. The dashed vertical line
represents the peak in the real part of the susceptibility. The inset shows
a slice of the FS through the $(001)$ plane, with the arrow depicting the
nesting that gives rise to the peak in $\mathfrak{Im}\;\chi_0({\bf q})$
between the hole (outer, red) and electron (inner, blue) FS sheets.}
\label{f:crchi}
\end{figure}

The energy shifts of the bands are also in good agreement with quantum
oscillations.  Comparing the semi-axis radii of the ellipsoids with high-quality
dHvA parameterizations (see Table \ref{t:vfs}), we find much improved
correspondence with experiment than the raw calculation, and indeed they compare
favorably with the shifts of Ref.\ \cite{major2004b}.  The differences in the
results of Ref.\ \cite{major2004b} and the current {\bf k}-space approach
reflect the different sensitivity of the two techniques to specific features of
the data (for example, compare the $N$-$P$ radius with the $N$-$H$ radius).  It
is also worth mentioning that the jungle-gym FS also originates from band 3, and
that this will also contribute to the shifts of this band, and so considering
the orbits about the $N$-ellipsoids alone may be misleading. Unfortunately,
there is a dearth of data for this sheet of FS, and comparisons are hard to
draw. As a final point, quantum oscillations appear to be relatively insensitive
to the FS of band 2, whereas we find a strong dependence of our fit to that
band, in agreement with other positron studies in V
\cite{singh1985b,pecora1988}.

Finally, for Cr (Fig.\ \ref{f:fitline}c), the data are reasonably well described
by the SD model throughout the BZ, whereas the IPM and JS models struggle near
the ${\Gamma}N$ points in both the raw band calculations and the rigid-band
fits.
In the absence of high-precision FS data for paramagnetic Cr, owing to the
ordering temperature ($T_{\rm N} \sim 312$K) of the spin-density wave, a robust comparison can instead
be made of the nesting vector of the paramagnetic FS that is widely believed to
determine the ordering (and has remained difficult to establish experimentally).
High-resolution neutron diffraction measurements have established the ordering
vector to be ${\bf Q} = (0, 0, 0.9516)\;(2\pi/a)$ (see Ref.\ \cite{fawcett1988}
and references therein). Our raw LMTO calculations predict (via a computation
of the static susceptibility, $\chi_0 ({\bf q})$, see for example
Ref.~\cite{laverock2009})
a nesting vector
${\bf q} \sim 0.930 \;(2\pi/{\bf a})$ (see Fig.\ \ref{f:crchi}),
which is rather smaller than the neutron measurements. Since quantum
oscillations are precluded in the paramagnetic phase (and have recently remained
unobserved from the relevant FS sheets in the ordered phase owing to strong
spin-fluctuation induced
scattering \cite{ditusa2010}), the only data on the FS
that has been capable of extracting this nesting vector have been some recent
angle-resolved photoemission measurements on Cr(110) thin films
\cite{schafer1999,rotenberg2005}, in which a nesting vector of ${\bf q} \sim
0.950 \pm 0.005\;(2\pi/{\bf a})$ is reported, in very good agreement with
neutron measurements. Our rigid-fit to the data (Fig.\ \ref{f:crchi})
culminates in a FS nesting vector of ${\bf q} \sim 0.950 \pm 0.002\;(2\pi/{\bf
a})$, where the error quoted is the combined error from shifting the two bands
to match the experimental results, representing the highest-precision
experimental confirmation of the relevant dimensions of the FS of paramagnetic
Cr from a bulk measurement. In contrast to this excellent agreement, the shifts
of the bands obtained using the IPM and JS models suggest nesting
vectors of ${\bf q} \sim 0.924 \pm 0.002\;(2\pi/{\bf a})$ and ${\bf q} \sim
0.932 \pm 0.002\;(2\pi/{\bf a})$ respectively.

The conclusions we draw from this section are the following. First,
our approach provides a robust empirical means of {\em measuring} the orbital
electron-positron enhancement factors, that are truly state-dependent (i.e.\
{\bf k}-dependent). Second, this measurement is not so strongly dependent on the
accuracy of the band calculation, being rather more sensitive to the overall
shape of the momentum distribution. Third, simultaneous fitting of the energy
bands {\em and} the orbital enhancement lead to a tuned FS that is in better
agreement with other FS data than is the raw band calculation, as well as
in good agreement with previous {\bf p}-space fitting approaches. Finally,
the band shifts that are required to reproduce experimental data (that is
in better agreement) are generally smaller for the SD model than the other
approaches investigated here, indicating that artificially large rigid shifts
in the bands can develop as a consequence of an inadequate description of
the enhancement.

\begin{table}[t]
\begin{center}
\begin{tabular}{||c||c|c|c|c||c||}
\hline
direction     & LMTO & Ref.\ \cite{parker1974} dHvA & SD fit & Ref.\
\cite{major2004b} fit \\
\hline
$N$-$P$       & 0.257 & 0.223 & $0.224\pm0.002$ & 0.245 \\
$N$-$\Gamma$  & 0.254 & 0.212 & $0.204\pm0.002$ & 0.231 \\
$N$-$H$       & 0.168 & 0.176 & $0.146\pm0.001$ & 0.160 \\
\hline
\end{tabular}
\end{center}
\caption{Comparison of the semi-axis radii (in units of $2\pi/a$)
of the $N$ hole ellipsoids from our
raw LMTO calculation and the fitted SD momentum density for V. Comparisons are
made with the high-precision parameterizations of dHvA data of Ref.\
\cite{parker1974} as well with the fitting technique (also applied to 2D-ACAR
data) employed by Ref.\ \cite{major2004b}. The errors reflect the error in
locating the minimum of the fit with respect to the shift in the bands. Note
that the dHvA radii rely on the assumption of perfect ellipsoids.}
\label{t:vfs}
\end{table}

\section{Simple metals}
\label{s:alkali}
We now turn to the other regime of enhancement, in which the bands, of $sp$
character, are closer to the nearly free electron model. Aluminium and the
alkali metals (and their alloys) provide a more stringent test for the SD
model. The electron-gas parameter of Al is $r_s = 2.65$, above the
point at which the JS (which does not conserve the low-density limit) and
the BN (which does) begin to diverge; as a consequence JS is not expected
to perform well here.

\begin{figure}[t]
\begin{center}
\includegraphics[width=1.0\linewidth]{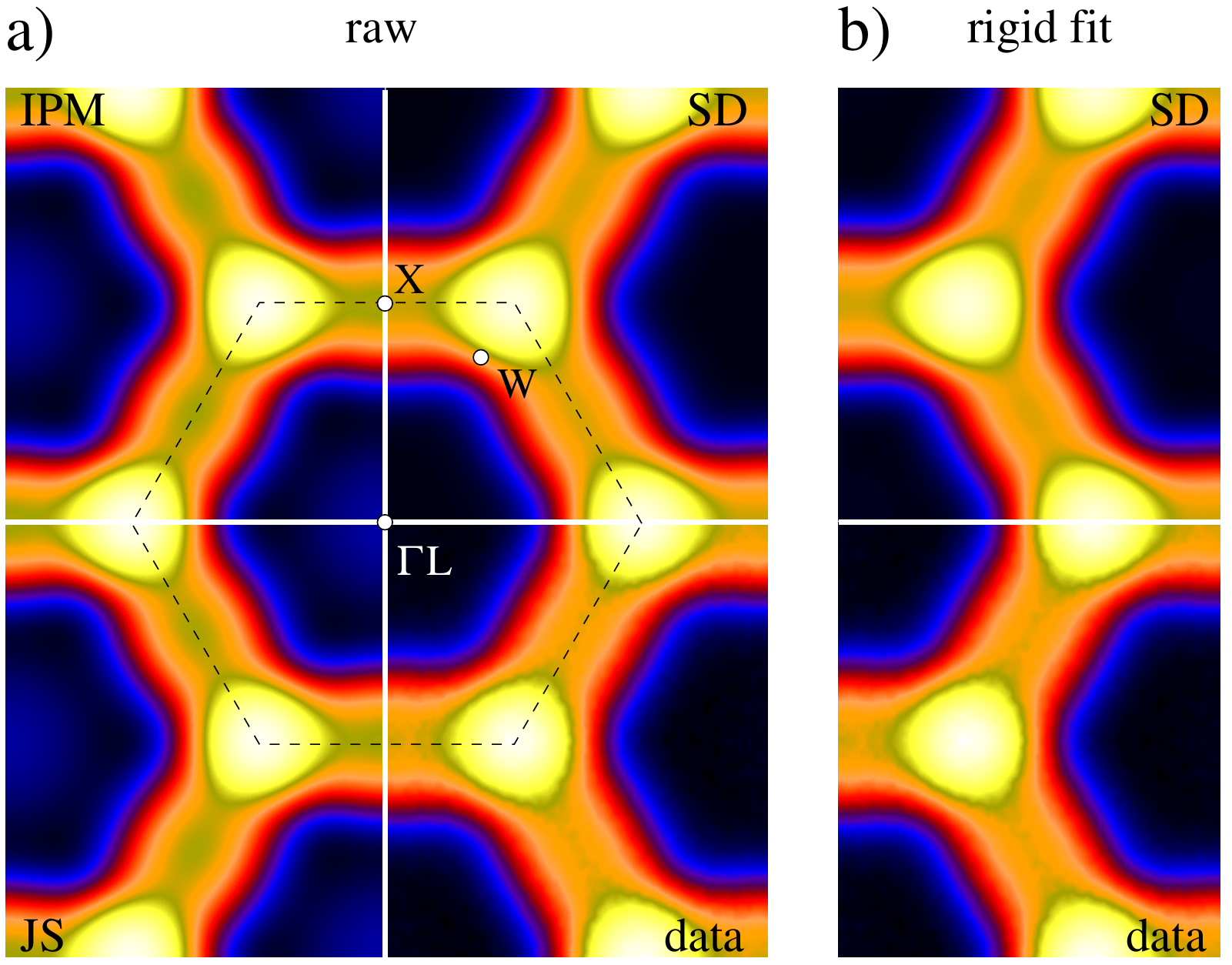}
\end{center}
\caption{(color online) Comparison between experimental data for Al projected
down the [111]
axis and the computed RMD for IPM, JS and SD models. (a) The raw LMTO band
calculation, and (b) the results of the rigid fitting of the energy bands. The
high symmetry points in projection ($\Gamma{L}$, $X$ and $W$) are shown in
(a), and the boundary of the first BZ is marked by the dotted line.}
\label{f:aldat}
\end{figure}

The FS of Al is composed of two bands, one which forms a $\Gamma$-centred hole
sheet from band 2 that lies completely in the first BZ, and from band 3 a
so-called `dismembered monster', that consists of square electron rings that run
the length of the edges of the first BZ except at the corners ($W$-points)
\cite{cracknell1969}.  Two 2D-ACAR projections along the [110] and [111]
directions were measured at room temperature and compared with LMTO calculations
performed over 1505 k-points in the irreducible BZ using the IPM, JS and SD
enhancement as before.

In Fig.\ \ref{f:aldat}, the data for the [111] projection is shown alongside the
LMTO calculations of the RMD. The FS structure can be clearly seen in the data
(shown in the bottom right panel), where the low density in the center reflects
the band 2 hole sheet, and the higher density at the edges of the projected BZ
come from the electron rings of band 3. At the corner of the projected BZ (near
the $W$-point), the particularly high region is due to the projection of the
rings in neighbouring zones along the $\left< 111 \right>$ directions. As can be
seen in the left panels of Fig.\ \ref{f:aldat}, the IPM and JS models are
particularly poor at describing the enhancement at the edges of the zone that
connect these strong features. Moreover, a small local peak at $\Gamma{L}$ that
is predicted by both IPM and JS is not observed at all in the data.
Quantitatively, as might be expected from the electron density of Al, the JS
model fairs poorly for the raw band calculation, and even worse than the IPM
(see Table \ref{t:alfit}). The SD model, however, does an excellent job of
describing the RMD of Al, correctly accounting for the absence of the local peak
at $\Gamma{L}$ and the connectivity of the strong features near $W$, and leading
to an almost order-of-magnitude improvement in the $\chi^2_{\rm red}$ parameter.
Here, the $s$-states are de-enhanced substantially, presumably owing to them
lying very low in energy. Unlike the previous $d$-electron systems, the
de-enhancement of the (unoccupied) $d$ states is not observed for Al.

\begin{table}[tb]
\begin{center}
\begin{tabular}{||c||cccc||c||}
\hline
 & $\gamma_s$ & $\gamma_p$ & $\gamma_d$ & $\gamma_f$ & $\chi_{\rm red}^2$ \\
\hline
\multicolumn{6}{||c||}{raw band} \\
\hline
  IPM&  -  &  -  &  -   &  -   & 15.40\\
  JS &  -  &  -  &  -   &  -   & 17.20\\
  SD & 0.53& 1.00&(1.29)&(0.60)&  2.29\\
\hline
\multicolumn{6}{||c||}{rigid fit} \\
\hline
  IPM&  -  &  -  &  -   &  -   &  3.93\\
  JS &  -  &  -  &  -   &  -   &  3.92\\
  SD & 0.60& 1.00&(1.09)&(0.72)&  2.12\\
\hline
\end{tabular}
\end{center}
\caption{The results of the fit between the different parameterizations of the
enhancement and the data for Al, shown in the same way as Table
\ref{t:rigidfit}.  The band-shifts that accompany the rigid-band fit are shown
in Table \ref{t:alshift}.}
\label{t:alfit}
\end{table}

\begin{table}[tb]
\begin{center}
\begin{tabular}{||c||cc||c||}
\hline
 & \multicolumn{2}{c||}{band shifts / mRy} & electron +/- \\
\hline
                                    & band 2 & band 3 &         \\
\hline
                                IPM & -39.2  & -30.0  &   +0.21 \\
                                JS  & -40.0  & -33.7  &   +0.22 \\
                                SD  & -11.9  & -10.2  &   +0.07 \\
\hline
\end{tabular}
\end{center}
\caption{The shifts in the energy bands for each of the rigid-band fits of Al.}
\label{t:alshift}
\end{table}

When the bands are fitted, the agreement between data and theory for each model
is very good. However, for the IPM and JS models the local peak at $\Gamma{L}$
persists, albeit at a much weaker amplitude. Moreover, consistent with the
previous conclusions, the shifts in the energy bands are substantially larger
for the IPM and JS models than the SD model (see Table \ref{t:alshift}),
leading to an electron excess of
$\sim 0.21$ (over a single FS sheet). For the SD model, this discrepancy is
much reduced, at just 0.07 electrons. Similarly to Ag, the FS of Al is already
well-described by the LMTO calculation, and comparisons with quantum oscillatory
data \cite{kamm1963,larson1967} agree with the raw band and SD rigid-band fit
to within ${\Delta}k_{\rm F} \sim 0.03\,(2\pi/a)$ ($\sim 15\%$ of the
resolution function).

\section{Phenomenological model}
\label{s:model}
Given the above results, we aim to find a phenomenological model that imparts
predictive capability on the calculation of the RMD. Taking the SD fitted
FS as a baseline,
we attempt to improve on the SBA model of the enhancement. The predictions
of the SBA enhancement are, in general, satisfactory, offering a similar
description of the experimental RMD (in some cases slightly better, in others
slightly worse) to the JS enhancement model. The predictions of the SBA model
can be understood largely from the perspective of the localization of the
states, in which $s$ and $p$ states experience similar enhancement over the IPM,
with the $s$ states in transition metals slightly less than $p$ due to their
slightly more localized nature in these systems. The $d$ states are enhanced
much less in the transition metals, associated with the greater localization,
and the increasing localization as the $d$-band becomes more filled is reflected
by the greater de-enhancement of the $d$-states in Ag when compared with
either Cr or V.

The Kahana model for enhancement, applied to a homogeneous electron gas and
parameterized in terms of $(k/k_{\rm F})^2$, is not expected to work
well for $d$-band systems, in which the effects of the crystal lattice can
completely hide the Kahana nature. For this reason, Mijnarends and Singru
(MS)
\cite{mijnarends1979} proposed a scheme parameterized by $\epsilon = (E-E_{\rm
bot})/(E_{\rm F}-E_{\rm bot})$, where $E_{\rm bot}$ is the energy at the bottom
of the conduction band,
\begin{equation}
\gamma = a + b \epsilon + c \epsilon^2,
\label{e:ms}
\end{equation}
where $a$, $b$ and $c$ are constants determined by the electron gas parameter
$r_s$. For a parabolic $s$ band this is identical to Kahana's formalism.
MS demonstrated the applicability of their prescription for the case of
Cu, in which substantial improvement was found (in {\bf p}-space) with this
description. The SBA model accounts for the variations in enhancement due to the
localization of a particular orbital, and its overlap with the positron
wavefunction, but does not consider the proximity of a state to the Fermi level,
leading to an over-estimation of the enhancement of more tightly-bound, filled
$s$ electron shells.
Adding such a scheme to the SBA model was not found to
universally explain the variations in enhancement for our experimental
data without different choices of the constants $a$, $b$ and $c$ (in fact,
following MS, we choose to set $a = 1$ in Eq.~\ref{e:ms} so that $b
\rightarrow b/a$ and $c \rightarrow c/a$).

\begin{table}[tb]
\begin{center}
\begin{tabular}{||c||c||c|c|c||}
\hline
\hspace*{0.3in} & \hspace*{0.1in} SBA \hspace*{0.1in} &
\multicolumn{3}{c||}{SBA-MS} \\
\hline
         & $\chi_{\rm red}^2$ & \hspace*{0.1in}$N(E_{\rm F})\hspace*{0.1in}$ &
         \hspace*{0.1in}$b/a$\hspace*{0.1in} & \hspace*{0.1in}$\chi_{\rm
         red}^2$\hspace*{0.1in} \\
\hline
V        &  20.41             & 23.80          & 0.700 & 4.30 \\
Cr       &   2.50             &  9.52          & 0.202 & 1.39 \\
Ag       &   4.71             &  3.60          & 0.042 & 4.50 \\
\hline
Al       &   7.90             &  5.04          & 0.589 & 2.28 \\
\hline
\end{tabular}
\end{center}
\caption{The linear component of MS-type energy-dependent enhancement ($b/a$)
obtained
by fitting the SBA model
to the data. $N(E_{\rm F})$ is given in units of states
/ Ry / unit cell, and the quadratic term, $c/a$, in Eq.~\ref{e:ms} is set
to 0. The $\chi_{\rm red}^2$ parameter is given before (SBA) and after (SBA-MS)
the application of the MS-type enhancement.}
\label{t:msfit}
\end{table}

Of particular interest in this analysis is the enhancement for V and Cr, which
are neighbors in the periodic table and would therefore, from the perspective
of a homogeneous electron gas, be expected to follow similar trends in their
enhancement owing to their similar electron density.  As can be seen in Table
\ref{t:rigidfit} the measured enhancement of V and Cr is quite different, and
yet V and Cr can each be well-approximated by a calculation of the other's
electronic band structure, with a simple extrapolation of $E_{\rm F}$ to account
for the different band-fillings (i.e. the rigid-band approximation works well).
The usual prescriptions for enhancement, in terms of the electron density, or
even a MS type energy-dependent enhancement, fail to predict such different
shell enhancements. Substituting the measured SD enhancement parameters for Cr
into the V calculation, and vice versa, is not found to describe the data well,
enforcing the idea that the enhancement is substantially and fundamentally
different for these two elements.
Since V and
Cr are electronically very similar, exhibiting the same body-centered cubic
structure, the largest difference between the two is in their band filling and
FS. In V, $E_{\rm F}$ lies close to a peak in the $d$ density of states
with appreciable
($\sim 20$ \%) $p$-character, leading to a total number of states at $E_{\rm F}$
of $N(E_{\rm F}) = 23.80$ states per Ry per atom. In paramagnetic Cr, on the
other hand, the additional electron places $E_{\rm F}$ in a valley between the
bonding and anti-bonding $d$-states with $N(E_{\rm F}) = 9.52$ states per Ry per
atom. It follows that the number of electrons that are capable of screening the
positron impurity (and thus lead to the enhancement of the annihilation rate) in
V and Cr is very different, and cannot be captured by considerations of the
electron density or energy alone. However, such a concept does provide a route
to understanding the different SD enhancement models in V and Cr, and the
different constants $b/a$ and $c/a$ in Eq.~\ref{e:ms} that are required to
explain the data.

In order to test such a correction to the enhancement, we apply a MS-type
enhancement to the SBA model, which is then fitted to the experimental
data. According to Kahana's theory, the quadratic part of the enhancement
parameterization is fairly weak, with $c/a \approx 0.138$ for metallic
densities, and can be well approximated by just a linear component
($b/a$). Here, we adopt just this linear energy enhancement,
and set $c/a = 0$ in Eq.~\ref{e:ms}, leaving just a single fitting
parameter that is capable of adjusting the {\em shape} of the computed RMD:
\begin{equation}
\gamma_{\rm SBA-MS} = \gamma_{\rm SBA} [1 + (b/a) \epsilon].
\end{equation}
The results of such a model, which we refer to as SBA-MS enhancement,
are found to enormously improve the agreement between data and theory
for all materials (see Table \ref{t:msfit}),
leading to $\chi_{\rm red}^2$ parameters that approach
the SD model investigated in Section \ref{s:dmetals}. Moreover, for the three
transition metal elements addressed in this manuscript, this linear component
of energy enhancement is found to scale with the density of states at the
Fermi level (Table \ref{t:msfit}), providing an empirical
model for the
enhancement of $d$-band elements and compounds. For $s$-$p$ electron metals,
the enhancement is found to more closely resemble Kahana's parameters, and
two regimes emerge -- Kahana's prediction for $s$-$p$ simple metals,
and a $N(E_{\rm F})$-dependent set of parameters for $d$ electron metals.
It is interesting to compare our results for an MS-type enhancement to those
applied by Matsumoto and Wakoh for Cr \cite{matsumoto1987}, in which they obtain
a factor $b/a \sim 0.15$ in a non-optimized approach, very close to our 0.20. On
the other hand, Genoud \cite{genoud1990}, employing an MS-type enhancement
independently for $s$, $p$ and $d$ electrons in V, obtained $b/a = 0.1 - 0.2$
for $s$ and $p$ electrons, also in a non-optimized way, which is somewhat
smaller than the optimum $b/a = 0.70$ that we find.

\begin{figure}[t]
\begin{center}
\includegraphics[width=1.0\linewidth]{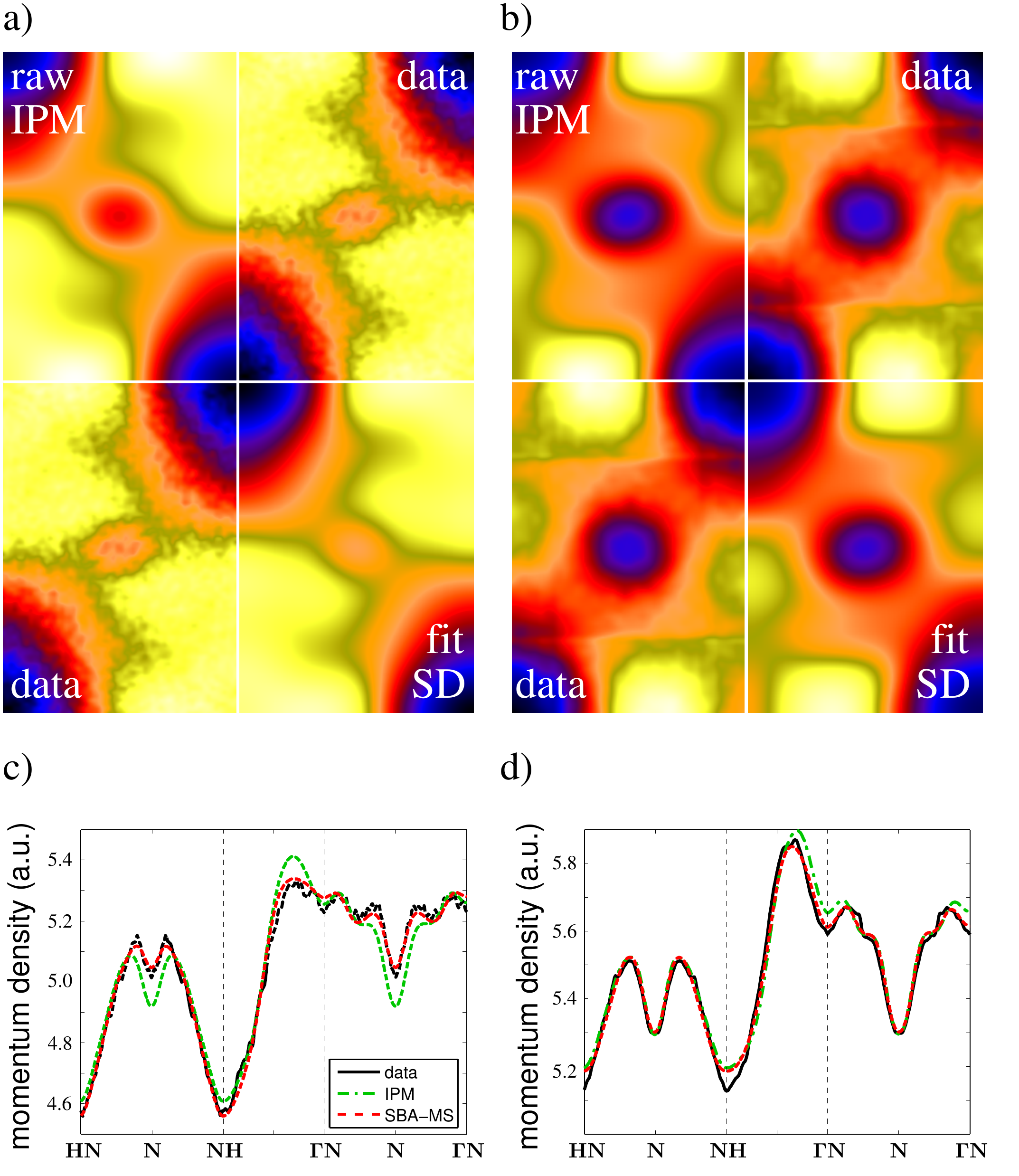}
\end{center}
\caption{(color online) Comparison between experimental data and the
corresponding theoretical
quantities in the raw IPM and rigid-band fit SD models of enhancement for (a) Cr
and (b) Mo. The RMD is shown along a path in the BZ in (c) and (d) for Cr and Mo
respectively.}
\label{f:crmo}
\end{figure}

Armed with such a model for the enhancement, we can now assess its
validity for another metal, specifically the $d$ electron metal Mo, for which we
have just a single [110] projection available, insufficient to permit a reliable
fit of the $\gamma_l$ parameters of the SD model. Instead, we apply
the SBA-MS model to the data, in which the $\gamma_l$'s are computed from
the partial annihilation rates of state $l$ and the energy-dependent enhancement
is provided from $N(E_{\rm F}) = 7.6$ states / Ry / atom and our preceding fit.
For comparison, we also compute the IPM and JS RMD. Both the JS and SBA models
of enhancement, by themselves, offer negligible improvement over the IPM, which
already provides a reasonable description of the data, and the application of
the SBA-MS model improves the agreement only modestly by $\sim 4$~\%. However,
lending freedom to the linear component $b/a$ is not found to provide any
additional improvement, emphasizing that the original IPM calculation was
already satisfactory.

One of the unresolved questions of 2D-ACAR in transition metals is why the {\bf
k}-space density of Cr and Mo (projected along the [110] direction) appear so
different, despite the apparent similarity of their isoelectronic and
isostructural FS topology \cite{dugdale1998} (see Fig.\ \ref{f:crmo}a,b).  In
Ref.\ \cite{dugdale1998}, maximum-entropy filtering techniques were employed to
assess the FS breaks in both distributions, ruling out the FS topology as an
explanation; the question of whether positron effects or consequences of the
proximity to magnetic structure in Cr are to blame were left open and have
remained so despite several efforts to resolve the issue, both experimentally
and theoretically \cite{dugdale2000,biasini2000,rubaszek2002}. Here, we are able
to solve this issue, which stems from a strong over-estimation of the
enhancement near $\Gamma{N}$ in Cr, previously highlighted in Fig.\
\ref{f:rawline}. In Fig.\ \ref{f:crmo}c,d, the RMD along a path in the BZ is
shown
for both Cr and Mo for the 2D-ACAR data, and the IPM and SBA-MS models of
enhancement. It is clear that the IPM (which resembles the JS model) looks
similar for both metals, eliminating positron wavefunction effects (which are
included in the IPM) as responsible for the strong difference in the data. On
the other hand, the SBA-MS prediction (which closely resembles our measured SD
model), accounts for the data very closely, unambiguously establishing
enhancement effects as the key.

\section{Conclusions}
We have presented a detailed investigation of the positron enhancement
factor for several metals, providing a quantitative {\em measurement} of the
state-dependence of the enhancement. By combining this with a rigid shift of
the energy bands, we demonstrate that, when the band structure is optimized to
2D-ACAR measurements, the precise location of the Fermi breaks in {\bf k}-space
are sensitively dependent on the accuracy of the enhancement model used in
the calculation. Furthermore, we show that, by employing a state-dependent
model for the enhancement, much improved agreement between the `tuned'
calculation and high-precision quantum oscillatory data can be obtained.
In particular, for Cr our positron measurements yield a nesting vector that
is in excellent agreement with neutron measurements of the spin-density wave
ordering vector, with an estimated accuracy better than 0.5\% of the BZ.
Although alloys have not been investigated here, this approach also allows
for the contribution from different atomic sites to be separated in the
experimental data, allowing a determination of the fraction of annihilations
from each individual element's hybridized wavefunctions (for example, see
Ref.~\cite{al3li}).

Comparisons of our (empirical) model with other popular models
of the enhancement have been made, particularly with the {\em ab
initio} state-dependent model of Barbiellini, Alatalo and co-workers
\cite{barbiellini1997}, for which a semi-empirical energy-dependent correction
is proposed that is found to bring the theory into much better agreement
with the data. Such a combined model therefore provides an accurate
model for the enhancement in momentum density measurements,
such as those of 2D-ACAR or coincidence Doppler broadening techniques.

\section*{Acknowledgments}
We acknowledge the financial support of the EPSRC (UK). We are indebted to the
late Maurizio Biasini for providing us with the 2D-ACAR data of Ag, and would
also like to thank Bernardo Barbiellini for stimulating discussions.

\end{document}